 \documentclass[sigconf]{acmart}

\AtBeginDocument{%
  \providecommand\BibTeX{{%
    \normalfont B\kern-0.5em{\scshape i\kern-0.25em b}\kern-0.8em\TeX}}}

\setcopyright{acmcopyright}
\copyrightyear{2025} 
\acmYear{2025} 
\setcopyright{rightsretained}
\acmConference[CHI EA '25]{Extended Abstracts of the CHI Conference on Human Factors in Computing Systems}{April 26-May 1, 2025}{Yokohama, Japan}
\acmBooktitle{Extended Abstracts of the CHI Conference on Human Factors in Computing Systems (CHI EA '25), April 26-May 1, 2025, Yokohama, Japan}\acmDOI{10.1145/3706599.3719678}
\acmISBN{979-8-4007-1395-8/2025/04}

\usepackage{graphicx}
\usepackage{wrapfig}
\usepackage[T1]{fontenc}
\usepackage{amsmath}  
\usepackage{multirow}
\usepackage{tabularx}
\newcommand{\x}[1]{{\leavevmode\color{black}{#1}}}
\usepackage{soul}
\acmSubmissionID{7302}

\begin{document}

\title{Imagining the Far East: Exploring Perceived Biases in AI-Generated Images of East Asian Women}
% Art as an Expression

\author{Xingyu Lan}
\affiliation{
  \institution{Fudan University}
  \city{Shanghai}
  \country{China}
  %\streetaddress{8600 Datapoint Drive}
  %\postcode{78229}
  }
\email{xingyulan96@gmail.com}
\authornote{Xingyu Lan is the corresponding author. She is a member of the Research Group of Computational and AI Communication at Institute for Global Communications and Integrated Media.}

\author{Jiaxi An}
\affiliation{
  \institution{Fudan University}
  \city{Shanghai}
  \country{China}
  }
\email{anjiaxi_lya@163.com}

\author{Yisu Guo}
\affiliation{
  \institution{Fudan University}
  \city{Shanghai}
  \country{China}
  }
\email{guoyisu2003@gmail.com}

\author{Chiyou Tong}
\affiliation{
  \institution{Fudan University}
  \city{Shanghai}
  \country{China}
  }
\email{t13661697442@gmail.com}

\author{Xintong Cai}
\affiliation{
  \institution{University of California, Los Angeles}
  \city{Los Angeles}
  \country{United States}
  }
\email{caixintong@g.ucla.edu}

\author{Jun Zhang}
\affiliation{
  \institution{Hubei Institute of Fine Arts}
  \city{Wuhan}
  \country{China}
  }
\email{alexmaya@tongji.edu.cn}

% macro
\newcommand{\etal}{et~al.~} 
\newcommand{\ie}{i.e.,~}
\newcommand{\eg}{e.g.,~}
\newcommand{\nbias}{18 }
\newcommand{\nimages}{300 }
\renewcommand{\shortauthors}{Lan et al.}

\renewcommand{\sectionautorefname}{Section}
\renewcommand{\subsectionautorefname}{Section}
\renewcommand{\subsubsectionautorefname}{Section}

\definecolor{myBlue}{HTML}{4E79A7}
\definecolor{myOrange}{HTML}{F28E2B}

\begin{abstract}
Image-generating AI, which allows users to create images from text, is increasingly used to produce visual content. Despite its advancements, cultural biases in AI-generated images have raised significant concerns. While much research has focused on issues within Western contexts, our study examines the perceived biases regarding the portrayal of East Asian women. In this exploratory study, we invited East Asian users to audit three popular models (DALL-E, Midjourney, Stable Diffusion) and identified \nbias specific perceived biases, categorized into four patterns: Westernization, overuse or misuse of cultural symbols, sexualization \& feminization, and racial stereotypes. This work highlights the potential challenges posed by AI models in portraying Eastern individuals.
\end{abstract}






%   \includegraphics[width=\textwidth]{figures/teaser.png}
%   \caption{Examples of AI-generated East Asian female images deemed biased by six participants in Study I (Sec. \ref{sec:audit}), compared with the ideal female images they provided.}
%   \Description{xxx}


\maketitle

\section{Introduction}

With the rise of generative artificial intelligence models (Generative AI), content creation is experiencing a revolutionary shift. Among the various forms of generative AI, image-generating AI is one that has matured earlier. In addition to its rapid image generation, image-generating AI has shown strong capabilities in comprehension, visual synthesis, style transfer, and the enhancement of human creativity~\cite{zhou2024generative}.
Today, image-generating AI has found fertile applications in industries such as art, news media, film production, gaming, and architecture. According to a report by Everypixel Journal in 2023~\cite{aistatistics}, AI generated 15 billion images within just 18 months, which is equivalent to the number of photos taken by human photographers over the past 150 years.
However, image-generating AI also exhibits concerning characteristics. For example, Ghosh and Caliskan~\cite{ghosh2023person} found that when requested to draw a ``person'', AI was more likely to draw a Western white male. 
% Sun~\etal~\cite{sun2024smiling} found that when generating images of professionals, DALL-E tended to reinforce rather than reduce gender stereotypes.
% (\eg ``housekeeping'' images mostly depicted females, while ``scientist'' images mostly depicted males). 
% Furthermore, compared to real census data and Google image search results, the gender ratios within these AI-generated images were even more imbalanced.
Cheong~\etal~\cite{cheong2023investigating} and Sun~\etal~\cite{sun2024smiling} discovered that AI models tend to reinforce both gender and racial stereotypes when depicting professions.
% most AI-generated images defaulted to portraying white individuals, with only a few professions showing a balanced racial distribution.
Such biases have also been reported by media outlets such as The Washington Post~\cite{wp} and Bloomberg~\cite{bloomberg}, sparking public attention.

\begin{figure}[h]
 \centering
 \includegraphics[width=\columnwidth]{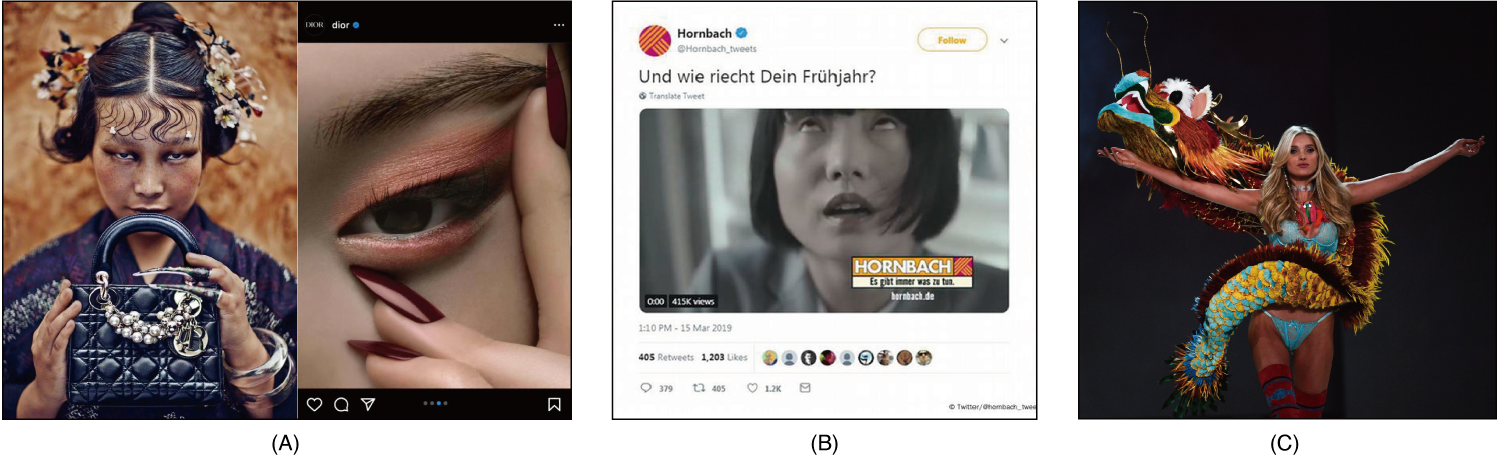}
 \caption{Examples of backlash against Western brands due to their portrayal of East Asian women: (A) Dior faced accusations of racism due to pulled eye advertisements and the uglification of East Asian women~\cite{diorads,diorads2}, (B) A German advertisement showing an East Asian woman getting aroused after smelling dirty laundry worn by white men fuelled anger in Asia~\cite{germanads}, (C) Victoria's Secret was repeatedly criticized for cultural appropriation~\cite{victoria}.}
 \Description{There are three images: (A) A Dior advertisement portrays East Asian women as fierce and with pulled eyes, (B) A German advertisement showing an East Asian woman becoming aroused after smelling dirty laundry worn by white men sparked anger in Asia, (C) At the Victoria's Secret show, a model walked the runway wearing a bright Chinese dragon costume, which was accused of cultural appropriation.}
 \label{fig:intro}
\end{figure}

However, while previous studies and discussions have laid a solid foundation in bias research, they have predominantly focused on Western contexts (\eg biases against African Americans) and recruited Western participants. This resonates with previous reflections within the HCI community, which noted that 73\% of study findings published at the leading HCI conference, ACM CHI, are based on Western samples, a demographic that represents less than 12\% of the world's population~\cite{linxen2021weird}. Thus, in this work, we seek to contribute an Eastern perspective by focusing on a specific theme: the portrayal of East Asian women in AI, a topic that is both classic in research (see \autoref{sec:related_1}) and a real-world concern.
% For example, Western film, fashion industries often depict East Asian females with slanted eyes as a representation of beauty, but this is frequently seen as racial discrimination in East Asia. 
For example, in recent years, Western brands such as Dior, Zara, and Dolce \& Gabbana have faced severe backlash and commercial value damage due to the inappropriate portrayal of East Asian women in visual media (\eg \autoref{fig:intro}). Although these brands sometimes explain that the portrayals are not driven by malicious bias, but rather by a pursuit of art or an attempt to highlight the uniqueness of East Asian women (\eg \cite{diorads2}), many East Asians indeed feel discomfort or offense due to complex historical, cultural, and political reasons. In the era of image-generating AI, we speculate that similar issues may also be triggered or exacerbated by AI-produced images.

However, it must be acknowledged that in such incidents, it is challenging to determine whether a controversial portrayal from the West results from intentional bias or merely reflects a lack of understanding of East Asian culture and differing aesthetics. Therefore, rather than distinguishing intentional biases from the producers' side, this work focuses more on the user side and specifically explores the \textbf{perceived biases} that East Asian users might report, thus providing insights for the development and refinement of related models and tools. Driven by this motivation, we selected three mainstream image-generating AI models (\ie DALL-E, Midjourney, Stable Diffusion) and invited East Asian users to perform algorithm audits focused on the theme of creating images of attractive East Asian women. As a result, we identified a total of \nbias perceived biases categorized into four patterns, including Westernization, overuse or misuse of cultural symbols, sexualization \& feminization, and classic racial stereotypes. 
\x{We did not examine images of males in this work, as their representation is shaped by distinct interpretive frameworks and origins, and the issue of women's images is currently more prominent in public discussion.
For example, historically, East Asian men have been depicted as loyal warriors, chivalrous swordsmen, and kung fu masters, but also as small and weak~\cite{azhar2021you,lu2013stressful}. A future analysis of how AI generates images of males would be a valuable addition to the field.}

\x{Positionality statement: The authors of this work were all born and raised in East Asia. This personal connection to the subject matter has motivated us to critically examine how AI-generated images may perpetuate or challenge existing biases and stereotypes of East Asian women. We recognize that our perspectives are influenced by our upbringing and cultural context, and we strive to approach our research with awareness and sensitivity to these influences. Our goal is to contribute to a more nuanced understanding of AI’s impact on representation within our cultural context.}
% Future research can expand this scope to include other Eastern nations and cultural groups, as well as to explore AI's portrayal of males. 

\section{Background and Related Work}

\subsection{East Asian Women in the Eyes of the West}
\label{sec:related_1}

``The West'' and ``The East'' are often used in pairs. Although they may seem to describe physical locations, the division between the East and the West is more about cultural perceptions \cite{mcneill1997we}. The term ``West'' is primarily based on Western European culture and extends to include regions such as North America and Australia. In contrast, ``East'' refers to regions not dominated by Western culture, with Asia being a representative example. This vast region can be further divided into distinct cultural groups, such as East Asia, South Asia, and the Middle East. Among them, East Asia (also called Far East), centered around China, Japan, and South Korea, has been deeply influenced by Confucianism, exhibiting cultural orientations distinct from the West \cite{lu2022surprising}. 
% For example, while Western cultures emphasize individual independence, East Asian cultures often view individuals as part of a collective \cite{markus1991culture,triandis1989self}. Western cultures promote confidence and self-motivation, whereas East Asian cultures place a higher value on humility and self-reflection \cite{heine2007search}. 

Regarding the appreciation of beauty, while there is a general consensus on aesthetics among humans (\eg symmetrical and harmonious faces~\cite{little2011facial}), there are still significant differences in the specific standards of attractiveness for women between the East and the West. For instance, traditional folk stories in East Asia often emphasize facial features over body shape when describing female beauty, with a preference for rounder faces and fairer skin \cite{cashin2024evolution}. Psychological research has indicated that East Asians have a lower sensitivity to sexual maturity when evaluating physical attractiveness compared to Western people \cite{cunningham1995their}.
Cultural and historical research has revealed that the portrayal of East Asian women has evolved alongside shifting power dynamics between the East and West \cite{zhou2006tianchao,spence1999chan}. In the early 14th century, the West saw China, the giant empire in the East, as a utopia, and greatly admired Chinese aesthetics.
% books such as \textit{The Travels of Marco Polo} introduced the prosperity and wealth of China to Europe, sparking a centuries-long fascination known as the Chinoiserie period. 
% During this period, the West admired and idealized Eastern aesthetics. Among upper-class women, wearing silks, embroideries, and garments adorned with Chinese motifs, as well as donning Chinese-style accessories, became a hallmark of fashion.
% China was seen as a utopia, and Chinese aesthetics were greatly admired by westerners.
% During this time, the West idealized the East as a utopia and heavily borrowed from Eastern aesthetics in clothing, textiles, and decorative arts~\cite{rovai2016luxury}. 
However, after 1750, with the rise of Western power, the image of East Asia plummeted~\cite{zhou2006tianchao}. 
% In the 19th century, the term \textit{yellow peril} was used by the West to describe East Asians, and their appearance was heavily criticized.
% For instance, British lawyer Sirr~\cite{sirr1849china} once described Chinese beauty in his book \textit{China and the Chinese} as: ``A dingy yellow complexion (overplastered with white cosmetic), high cheek bones, small piggish-looking-eyes, with pencilled eyebrows meeting over the nose, low brow, oblong ears, coarse black hair, which is invariably anointed with stinking pork fat, until it stands on end.'' 
Stereotypes such as ``piggish eyes'' and ``slanted eyes'', along with perceptions of Eastern women as exotic, quiet, passive, and controllable, later permeated Western films, comics, and the fashion industry~\cite{marchetti1994romance,xing1998asian,azhar2021you}. 
% Such derogatory portrayals of East Asian women, including their stereotypical appearances and views of them as exotic, passive, and controllable, later became common in Western media such as fictions, films and ads.~\cite{yamamoto2000visible,uchida1998orientalization}, constituting the traumatic memories of racial discrimination perceived in the East Asian region today (\eg cases in \autoref{fig:intro}).
% For example, the narrative of a heroic White man and a fragile Asian girl (also called China Doll or Geisha Girl) is a classic archetype~\cite{yamamoto2000visible}. \textit{Madama Butterfly}, for example, portrays an American naval officer named Pinkerton who marries and then abandons a 15-year-old Japanese geisha named Cio-Cio-San. Cio-Cio-San waits for him her entire life and ultimately commits suicide.
% Another archetype is the Dragon Lady, which portrays East Asian women as mysterious and enigmatic~\cite{yamamoto2000visible}. In \textit{The Red Lantern}, one of the earliest Hollywood portrayals of a Chinese female, the character is depicted as young and alluring, serving as the mistress of a British white man.
% 20世纪的好莱坞，华裔女性出演的大多是情妇、奴隶等微末角色
% Similarly, the image of Japanese geishas has frequently been used by the West to depict East Asian women as quiet, submissive porcelain dolls~\cite{uchida1998orientalization}, .
However, traditional research on the imagery of East Asian women has predominantly been conducted within the domains of literature, film, history, and political science. Therefore, this study seeks to re-examine this classic subject within the contemporary context of generative AI. 
% By employing both qualitative and quantitative methods, we examine the perceived biases regarding East Asian women portrayed by three prevalent Western image-generating AI models.

% In brief, previous research on East-West facial prejudice has primarily approached from perspectives of aesthetics, literature, anthropology, and political science, exploring the origins and reasons for these differences mainly through historical research and textual analysis. This study, however, aims to look towards the future, uncovering new expressions and changes of this issue in the digital era. Specifically, we will focus on the rapidly evolving field of generative AI, which has significant social impact yet appears neutral under the guise of technology, to empirically investigate whether and in what form aesthetic prejudices against East Asian women exist in mainstream image-generating AI models.

\subsection{AI Biases}
\label{sec:related_2}

With the advent of AI, the imperative to develop trustworthy and responsible AI systems and to minimize AI-related harm has become increasingly pressing~\cite{shelby2023sociotechnical,courtland2018bias,kaur2022trustworthy,kidd2023ai}. AI bias, in particular, has emerged as a significant concern~\cite{ulnicane2023power}. 
% Huang~\etal~\cite{huang2022overview} considered fairness and justice as one important dimension to evaluate the ethics of AI.
% The causes of bias are complex, stemming from the training data itself as well as imperfect human logic in model construction and evaluation~\cite{kaur2022trustworthy}.
% However, they are not infallible; they might introduce bias into a system due to their probabilistic nature, imperfect human logic used in their development, or non-representative input or training data from biased individuals (Seaver, 2013; Diakopoulos, 2014). 
To date, while various technologies have been developed to identify and mitigate bias~\cite{roselli2019managing}, those that are more easily addressed often pertain to the more objective \textit{technical biases}. In contrast, \textit{social biases}, which are relevant to human concerns, are more complex, controversial, and challenging to tackle~\cite{ulnicane2023power}.
For example, researchers have detected price discrimination targeting specific regions and social classes on platforms such as Google, Bing, and Amazon~\cite{mikians2012detecting,mikians2013crowd}. 
\x{Besides, a set of algorithmic auditing endeavors have shown that women face varying degrees of disadvantages in algorithmic environments~\cite{hall2023systematic}, such as being assigned less diverse and more stereotypical labels~\cite{papakyriakopoulos2023beyond,noble2018algorithms}, being less likely than men to be recommended for high-paying positions and roles in the STEM fields~\cite{lambrecht2019algorithmic,bogen2018help}, and being overclassified as high risk in criminal risk assessments~\cite{hamilton2019sexist}.}
% groupings such as Google have been found to exhibit gender and racial biases~\cite{noble2018algorithms,diaz2008through,papakyriakopoulos2023beyond}. 
% In the realm of algorithmic job recommendations, women have been found to be . 
% Researchers have also found algorithmic biases in fields such as children's welfare, medical diagnoses, political elections, and business evaluations~\cite{eslami2017careful,obermeyer2019dissecting,urman2022matter,chouldechova2018case}. 
% These AI bias issues have also garnered widespread public attention. For instance, 
% in 2016, ProPublica revealed that algorithms used to predict crime risk in the US were more likely to classify Black individuals as high-risk~\cite{machine_bias}, leading to a series of heated debates and discussions~\cite{flores2016false,crawford2021atlas}.

With the rise of image-generating AI, biases have also been detected in visual forms. For example, Ghosh and Caliskan~\cite{ghosh2023person} found that Stable Diffusion was more likely to generate images of white males when prompted with the term ``person''. Cheong~\etal~\cite{cheong2023investigating} discovered that DALL-E defaulted to portraying white individuals in occupational images, with only a few professions, such as rapper, featuring more Black individuals. Sun~\etal~\cite{sun2024smiling} found that AI exhibits not only \textit{representation bias} (\eg depicting ``housekeeping'' as female and ``scientist'' as male) but also \textit{presentation bias}, such as portraying women more often as happy, smiling, and in submissive positions.
\x{Mack~\etal~\cite{mack2024they} found that AI often oversimplifies the depiction of people with disabilities and displays negative stereotypes (\eg being in a wheelchair or appearing helpless).}
Karpouzis~\cite{karpouzis2024plato} drew on Plato's \textit{Allegory of the Cave} and commented that cultural biases in AI could exacerbate the echo chamber effect, where mainstream culture is amplified while minority voices are weakened or silenced.
% Compared to traditional AI bias research, which mostly focuses on search engines and recommendation systems, image-generating AI has shown a set of new features. For example, traditional AI mediums retrieve and recommend content based on existing data without generating new content, whereas image-generating AI can produce large volumes of new images, disrupting the existing image market. Besides, image-generating AI is more interactive, creating content through user interactions (\eg write, draw, speak). Kidd and Birhane~\cite{kidd2023ai} argued in \textit{Science} that users may be more inclined to ``perceive generative models as knowledgeable, intentional agents implies a readiness to adopt the information that they provide more rapidly and with greater certainty''. 
Hence, there is a pressing need for more empirical research to deepen our understanding of image-generating AI. 
% However, so far, research in this area remains limited, with existing studies primarily focusing on \textit{representation biases} (what is depicted) and giving less attention to \textit{presentation biases} (how it is depicted). Therefore, this work aims to audit the images of East Asian women portrayed by AI, offering insights from an Eastern perspective and uncovering more findings about presentation biases.

\section{Exploring Perceived Biases in the Portrayal of East Asian Women}
\label{sec:audit}

In this section, we report an exploratory study that examines whether and how users perceive biases regarding East Asian women drawn by image-generating AI models.

\subsection{Methodology}

\subsubsection{Algorithm audit}
\label{sssec:auditdetail}

Algorithm audit is an empirical research method used to investigate potential risks in algorithmic systems, particularly the manifestation of social problems~\cite{bandy2021problematic}. 
% Typical methods of algorithmic audit include code audit, noninvasive user audit, scraping audit, sock puppet audit, and collaborative or crowdsourced audit~\cite{sandvig2014auditing}. Each of these methods has its advantages and disadvantages, depending on the research question. For example, a code audit directly examines the source code, enabling precise identification of algorithmic issues but requiring access to the actual source code to understand the algorithm's operational logic. A scraping audit is more suitable for analyzing algorithmically generated content stored and presented on specific platforms (\eg websites, apps). 
A crowdsourced audit is a user-driven method~\cite{devos2022toward} that recruits real-world users to audit algorithms and gather feedback to uncover algorithmic issues. This method is particularly suitable for exploratory studies.
Therefore, we finally chose to invite East Asian users to audit three mainstream image-generating AI models: (i) DALL-E 3 model by OpenAI (USA), (ii) Midjourney v6 model by Midjourney Lab (USA), and (iii) Stable Diffusion 3 model by Stability AI (UK). These models represent the most widely used and highest-output text-to-image AI models~\cite{aistatistics}.

\subsubsection{Participants}

We recruited participants by posting advertisements on social media and snowballing. The recruitment materials specified the following requirements: (i) Participants should come from East Asian countries (China, Japan, South Korea), predominantly reside within East Asia, and self-identify as East Asian. (ii) Participants should have knowledge and some experience with image-generating AI. (iii) Participants should be proficient in English to operate English interfaces.
% All materials and procedures described below were approved by the institutional review board at the authors’ university. 
A total of 36 qualified participants were recruited (25 females, 11 males), including 20 Chinese, 7 Japanese, and 9 Koreans. Their ages ranged from 18 to 50 (\textit{M} = 23, \textit{SD} = 6.3). Since the study required the participants to navigate English interfaces and write prompts in English, all participants had college education (Bachelor's Degree: 88.89\%, Master's Degree: 11.11\%). Their educational backgrounds were diverse, encompassing fields such as social sciences, engineering, medicine, and management. The study was conducted offline in China with each participant individually. All participants received a compensation of \$8.

\subsubsection{Study Procedure}

To start with, we introduced the study's background and provided a tutorial on DALL-E, Midjourney, and Stable Diffusion. After the participants became familiar with these tools, they were tasked with ``drawing an attractive East Asian woman'' by exploring and using each tool to perform the task. 
\x{We used the term, ``attractive'' (\ie a person’s physical features are considered aesthetically pleasing or beautiful~\cite{liu2020encyclopedia}), following the established research tradition in psychology that has extensively studied physical attractiveness since the 20th century (\eg ~\cite{walster1966importance,berscheid1971physical,cunningham1995their,little2011facial}). Compared to words such as ``beautiful'' and ``good-looking'', this term not only encompasses multiple dimensions, such as facial features, body proportions, and skin condition, but also raises fewer ethical concerns, as it is less likely to be perceived as judgmental or discriminatory.}
For each generated image, they were asked to record the image, the tool used, and the prompt in a spreadsheet. To help the participants get started, we provided some pre-prepared keywords, including (i) adjectives such as ``attractive'', ``beautiful'', and ``pretty''~\footnote{Mainstream image-generating AI platforms generally have ethical checks in place. Our pre-study test showed that if users input insulting or offensive terms, such as ``ugly'', the platforms either refuse to create the image or convey negativity by only adjusting clothing or distorting facial features. Therefore, in this study, prompts used only unidirectional adjectives to ensure a focus on creating attractive women.}; (ii) adverbs describing intensity, such as ``very'' and ``highly''; (iii) nouns relating to women, such as ``female'' and ``lady''; (iv) nouns indicating regions, such as ``East Asia'', ``China'', ``Japan'', and ``Korea''. These keywords were merely provided for inspiration, and participants were encouraged to write their own prompts as they wished. We also provided participants with an English dictionary and a thesaurus tool to help them write their own prompts. 
% To capture as many details as possible, we followed previous research~\cite{devos2022toward} and encouraged participants to use the think-aloud method, voicing their real-time thoughts and feelings. Each participant spent approximately one hour on the drawing task.

Next, we conducted interviews with each participant, focusing on questions such as: ``What should be the characteristics of an attractive East Asian woman?'', ``Could you provide a photo representing your ideal woman image?''.
% ``Do you think the women generated by the AI align with your expectations?'' 
Besides, we asked the participants to go through all the images they generated one by one, explaining whether they perceived any biases, and why.
% We also posed targeted questions based on the participants' verbalized reactions (\eg ``You expressed surprise earlier, could you please elaborate on what happened at that moment?''). 
Lastly, the participants rated the three models in terms of their overall satisfaction on a scale from 0 to 100. The interviews were audio recorded with the participants' consent.

\subsection{Analyses and Findings}

We collected 675 AI generated images from the study (\ie 18.75 images per participant on average) and 36 images of ideal East Asian women provided by our participants. 
We also obtained an abundance of qualitative data elucidating the participants' opinions and explanations. Below, we report our data analyses and major findings. Note that when quoting participants, we use the format P1-P36 to refer to their IDs, with subscripts indicating their nationality (CN: China, JP: Japan, KR: Korea) and the tool they commented on in the quote (DE: DALL-E; MJ: Midjourney; SD: Stable Diffusion). For example, $P1_{CN,DE}$ refers to Participant 1, who is Chinese and commented on DALL-E.

% \vspace{-1em}
\begin{figure*}[t!]
 \centering
 \includegraphics[width=\textwidth]{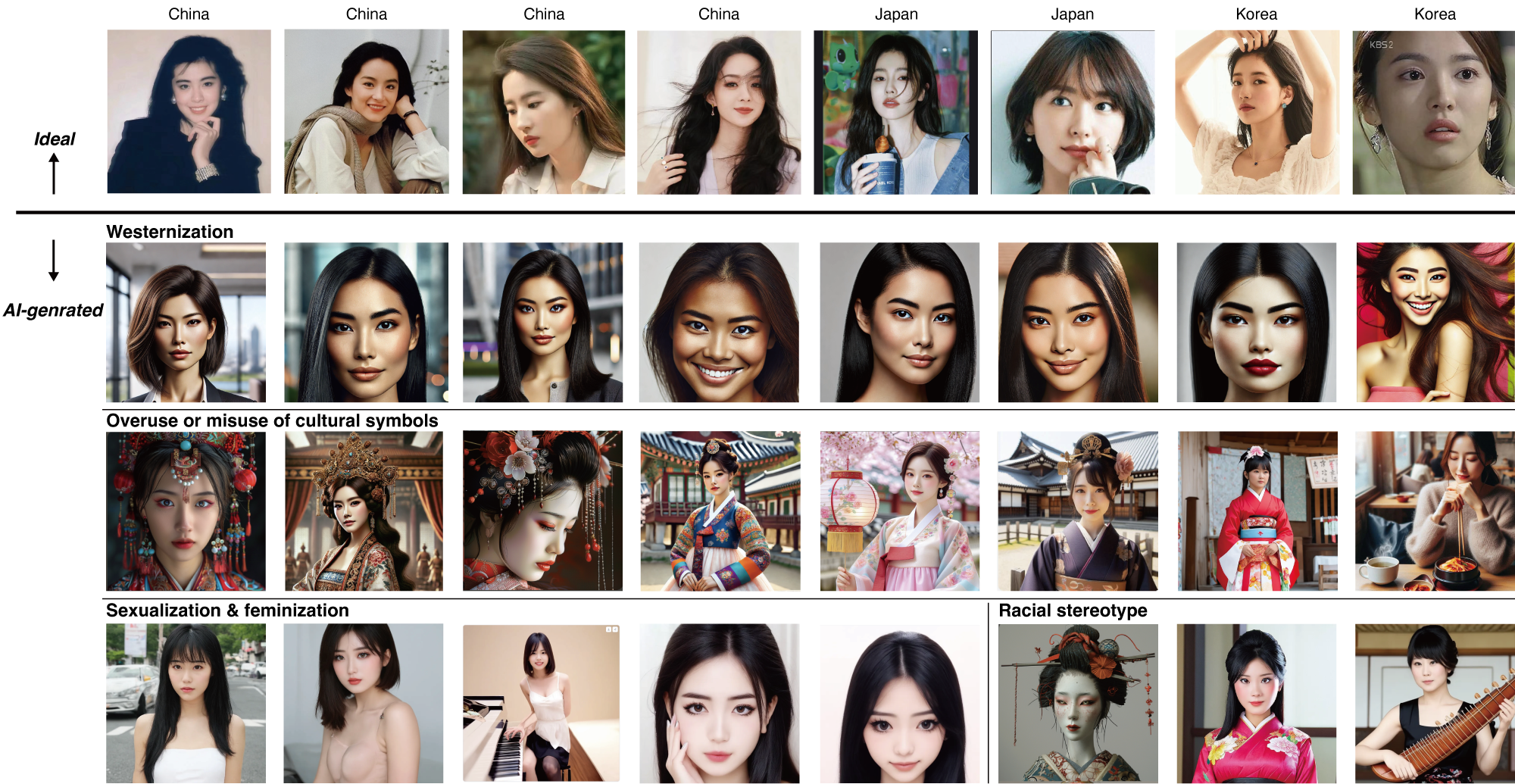}
 \caption{Examples of ideal images of East Asian women provided by our participants, and examples of AI-generated images.}
 \label{fig:examples}
  \Description{The first row displays eight images of ideal East Asian women as provided by our participants. The second row showcases eight images that participants reported as exhibiting Westernization biases. The third row presents eight images where participants noted an overuse or misuse of cultural symbols. The fourth row displays five images that participants identified as reflecting biases related to sexualization and feminization, along with three images featuring the racial stereotype of slanted eyes.}
\end{figure*}

\subsubsection{Ideal images}
\label{sssec:ideal}

We analyzed the participants' qualitative descriptions of their ideal East Asian woman images using natural language processing techniques. The most frequently mentioned nouns associated with ideal women were \textit{eyes} (N = 16), \textit{facial features/face} (16), \textit{temperament} (9), \textit{style} (6), and \textit{hair} (6).
The most mentioned adjectives were \textit{overall (feeling)} (8), \textit{soft/gentle} (8), \textit{fair (skin)} (7), \textit{well-proportioned/harmonious} (6), \textit{natural} (4), and \textit{healthy} (4).
Overall, the participants' ideal images of East Asian women were highly consistent with the indigenous beauty standards identified in previous literature~\cite{cashin2024evolution,han2003body,jung2006cross,dimitrov2023beauty}, which emphasize facial features (particularly the eyes) over body shape, and prefer roundness over sharpness.
For example, our participants described their ideal image of a woman as``\textit{having bright eyes}'' ($P5_{CN}$); the eyes ``\textit{don't need to be very large}'' ($P34_{KR}$) and ``\textit{should be luminous}'' ($P3_{CN}$). There was also a pursuit of fair skin (\eg $P21_{JP}$: ``\textit{a bit whiter than yellow}'', $P27_{CN}$: ``\textit{skin should be fair but not as white as Westerners}''), as well as a pursuit of harmonious and round facial features (\eg $P13_{CN}$: ``\textit{no sharp angles}'', $P32_{KR}$: ``\textit{overall balance is more important than the beauty of each individual facial feature.}'') 
% ``not seeking a particular characteristic, just pleasing to the eye'' (P11), 
Echoing previous finding~\cite{cunningham1995their}, our participants were not sensitive to features associated with sexual maturity. None of them mentioned keywords such as ``sexy''.
Most participants preferred a softer and elegant temperament (\eg $P28_{JP}$: ``\textit{not too fierce, more gentle}'', $P19_{CN}$: ``\textit{elegant and poised, appearing very dignified and graceful}'', $P35_{KR}$: ``\textit{I like women who are polite and cultured}''). 
% Among all the participants, only $P5_{CN}$ stated that her ideal woman should be ``\textit{wild and ambitious}''. 
% and P11 emphasized a preference for ``natural strength''
Besides, the participants generally favored natural beauty, and five of them had intentionally emphasized the avoidance of heavy makeup (\eg $P4_{CN}$: ``\textit{neat and natural, absolutely no heavy makeup}'', $P24_{JP}$: ``\textit{looking natural, without complicated makeup}''). 
% P21: ``a bit purer in appearance, not too three-dimensional'', P22: Korean: ``small face, agile appearance'', P15: ``more reserved, with charm''
% P13: ``beauty that is strong yet soft, elegant, calm, and dignified''
% ``should be real (something seen in real life), not too fake'' (P17)

However, many AI-generated images deviated from their ideals. Only 43.7\% of the generated images were reported as satisfactory by our participants (49.6\% of DALL-E images, 52.7\% of Midjourney images, and 14.4\% of Stable Diffusion images). 
% while 54.9\% were considered unsatisfactory
Participants frequently had to fine-tune the AI through multiple rounds to obtain a relatively satisfactory image. None of them received a satisfactory image right from the start.

\subsubsection{Taxonomy of Perceived Biases}
\label{sssec:taxonomy}

Next, we analyzed whether participants reported any perceived biases in AI-generated images with two research questions in mind: (i) \textit{where} the biases occurred, and (ii) \textit{how} the participants interpreted the biases. We adopted the open coding method. Two coders first independently performed their coding.
For (i), we generated codes based on the specific features that the participants identified as biased. For example, some participants reported perceived biases regarding \textit{lips} and \textit{skin}. For (ii), we coded participants' explanations and justifications of the biases, revealing why they perceived a certain feature as biased. For example, when reporting their unease about thick lips and heavy makeup, the participants often provided interpretations such as ``\textit{It suddenly reminds me of those Asian characters seen in American dramas}'' and ``\textit{It looks more like a Western style}''. Thus, we grouped these codes into a higher-level category called \textit{Westernization} to indicate their pattern of bias.
% \x{Additionally, no participants interpreted revealing clothing as Western cultures - but uncomfortable ningshi of women. }
% revealing clothing - passiveness, sexualized
Next, the two coders met to compare their codes, unify the naming of similar codes, and discuss any inconsistencies. 
% For example, we initially had two codes related to eye shapes: \textit{slanted eyes} and \textit{upturned eye makeup}. However, during discussions, we reached a consensus that these two aspects achieve a similar effect and are closely intertwined in public controversies (\eg ~\autoref{fig:intro} A). Therefore, we consolidated them into a single code named \textit{slanted eyes}. 
% and updated our codebook to specify that this code encompasses both the naturally upturned eye shape and the makeup that enhances this appearance
As shown in ~\autoref{tab:taxonomy}, \nbias reported biases were identified, categorized into four patterns. Examples of these patterns are shown in \autoref{fig:examples}.
% Below, we introduce each pattern and its corresponding biases in detail.
% The biases include not only facial features but also body-related and adornment-related features (tagged with icons in ~\autoref{tab:taxonomy}). We grouped the biases into 

\begin{table*}[t!]
\centering
\small
\fontsize{7.5pt}{10pt}\selectfont
%\begin{adjustbox}{width=\textwidth}\begin{tabularx}
\begin{tabularx}{\textwidth}{ p{8.5em}|p{12em}p{2em}X }
\toprule
\textbf{Bias patterns} & \textbf{Issues} & \textbf{Models} & \textbf{Description} \\
\midrule
Westernization (144)  &  western-styled makeup (27)	&	\includegraphics[height=0.7em]{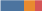} &  thick eyelashes and eyeshadow, broad eyebrow shapes, and heavy blush 	\\
& bony/sharp face (24)	&	\includegraphics[height=0.7em]{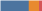} & Salient bone structure, such as high cheekbones and a pointed chin 	\\
& eyes (22)	&	\includegraphics[height=0.7em]{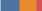} & Deep eyes, long eye fissures, double eyelids, short intercanthal distances 	\\
& nose (21)	& \includegraphics[height=0.7em]{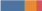} &	A high nose, narrow nasal widths 	\\
& lips (21)	&	\includegraphics[height=0.7em]{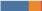}&  Full lips (or using lipsticks to emphasize full lips)	\\
& skin tone (15)	& \includegraphics[height=0.7em]{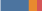} & Skin color leans towards the tanned wheat and chocolate shades 	\\
& big smile (14)	& \includegraphics[height=0.7em]{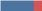} &	A broad smile showing teeth, with smile muscles engaged 	\\
% & chin (11)	& \includegraphics[height=0.7em]{figures/dist.png} &	A pointed and prominent chin (with straight and chiseled cheeks) 	\\ 
\midrule

Sexualization \& feminization (64)  & revealing clothing (15)	&	\includegraphics[height=0.7em]{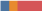} & Clothing that reveals breasts or buttocks, such as low-cut tops, deep V-necks, and mini skirts	\\
 & porcelain skin (12)	& \includegraphics[height=0.7em]{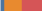} &	Highly smooth and flawless skin, lacking wrinkles and texture	\\
 & youth (10)	&	\includegraphics[height=0.7em]{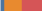} & Drawing young women by default; difficult to portray beautiful middle-aged or elderly women  	\\
& dull eyes (9)	& \includegraphics[height=0.7em]{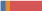} &	The gaze is dull, lacking confidence, avoiding direct eye contact	\\
& harsh gaze (9)	& \includegraphics[height=0.7em]{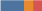} &	The gaze is fierce, piercing, and dangerous	\\
& slenderness (5)	& \includegraphics[height=0.7em]{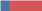} &	Drawing a slender figure by default; difficult to portray plump beauties 	\\
& long hair (4)	& \includegraphics[height=0.7em]{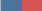} &	Drawing long hair that extends beyond the shoulders by default	\\ \midrule

Overuse/misuse of cultural symbols (63) 	& ancient/exotic symbols (25) & \includegraphics[height=0.7em]{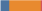} & Placing women in exotic backgrounds or dressing them in traditional clothing, jewelry, and patterns 	\\
 & mix of cultural elements (22)	&	 \includegraphics[height=0.7em]{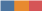} & Inability to distinguish between the differences within East Asian countries, leading to mixed use of elements 	\\
& wrong semantics (16)	& \includegraphics[height=0.7em]{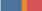} & 	Not in line with the real environment and living habits of East Asia	\\ \midrule

Racial stereotype (13) & slanted eyes (13)	& \includegraphics[height=0.7em]{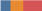} & 	Eyes that are narrow in width and angled upward at the outer corners (or makeup that emphasizes this trait). 	\\
% bony face
\bottomrule
\end{tabularx}
% \end{adjustbox}
\caption{A taxonomy of the perceived biases identified in our study. The ``Models'' column shows the relative distribution of these biases across the three models. \includegraphics[height=0.7em]{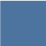} : DALL-E, \includegraphics[height=0.7em]{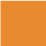} : Midjourney, \includegraphics[height=0.7em]
{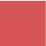} : Stable Diffusion.}
\Description{A taxonomy of the perceived biases identified in our study. The ``Models'' column shows the relative distribution of these biases across the three models using 100\% stacked bar charts. Blue: DALL-E, Orange: Midjourney, Red: Stable Diffusion.}
\label{tab:taxonomy}
\end{table*}

\textbf{Westernization.}
Westernization refers to the portrayal of East Asian women in ways that cater to Western physical appearance, styles, and values (see \autoref{fig:examples}, the second row). Within this pattern, the most mentioned bias is the use of Western-style makeup. In these images, although the women have East Asian faces, the makeup is applied to align their appearance more closely with Western preferences for bone structure and sexual maturity~\cite{cunningham1995their}. This practice deviates from the East Asian aesthetic preference for minimal or natural makeup (as reported in \autoref{sssec:ideal}).
For example, $P22_{KR,DE}$ said, ``\textit{The eye makeup is very deep and heavy; I don't think ordinary Koreans wear makeup like this.}'' $P11_{CN,DE}$ complained that ``\textit{I just feel that her eyebrows are too thick.}'' She then tried to fine-tune the prompts and found that ``\textit{there wasn’t any substantial improvement}.'' $P21_{JP,MJ}$ also commented that ``\textit{the makeup it produces feels very foreign for me.}'' 
% \w{$P?_{CN,DE}$ complemented, ``\textit{In East Asian social etiquette, heavy makeup is often considered inappropriate, overly showy, and even frivolous. This (the image) does not align with the image of a woman who is cultured and sensible.}} 
% This does not align with the image of a woman who is cultured and sensible.
Apart from makeup, our participants also reported instances where AI-generated images drew East Asian women with Western physical traits, such as eyes, noses, lips, and bony/sharp faces. For example, $P22_{KR,MJ}$ commented, ``\textit{The facial features are too Western; the eyes seem too large.}'' $P10_{CN,DE}$ noted that the image she received ``\textit{looks like an American-born Asian, and the cheekbones remind me of Kim Kardashian.}''
Such blending of Western features with East Asian facial structures often results in what is termed a pan-Asian face, which typically suggests a White ancestor in the person's family tree~\cite{yip2019beyond}. In Asia, this phenomenon is more common in countries with a history of significant Western colonization and racial mixing, such as those in South and Southeast Asia\x{~\cite{riddick2006history,lockard2009southeast}}. Therefore, we also observed complaints such as, ``\textit{Although the girl is also beautiful, her appearance clearly has a Southeast Asian style}'' ( $P3_{CN,SD}$) and ``\textit{The face shape looks like that of a foreigner, specifically a Thai person}'' ($P25_{JP,DE}$). $P4_{CN,DE}$ said, ``\textit{Look at the skin color, it’s not East Asian; it's very dark, like a mixed-race with Black and White ancestry.}''
% $P17_{CN,DE}$ said, ``\textit{(the woman) looks not like East Asian, she has a sense of mixed-race.}''
% 【p4-9-4】这一些我觉得就是给我一种要说印尼亚人也好，或者那种 ABC 也好……（高颧骨）（dalle）
% Additionally, we also observed two features that the participants thought reflect Western values and social norms, including the skin color and big smile. For example, 
Additionally, several participants reported perceived bias related to the depiction of a big smile. For example, $P3_{CN,SD}$ noted that showing a lot of teeth is not a common or ideal expression for East Asian women, but the model ``\textit{always produces a very exaggerated big smile.}'' 
$P6_{CN,DE}$ attributed such portrayals to Western culture, which emphasizes confidence: ``\textit{The smiling muscles are too intense and exaggerated, resembling that of a very confident and sunny ABC (American-born Chinese).}''

% It should also be noted that not all participants perceived biases uniformly due to westernization. For example, $P24_{JP,DE}$ thought a mixed-race face  is ``\textit{acceptable for me.}'' $P10_{CN,DE}$ said the AI-generated image ``\textit{matches my imagination of a beautiful ABC lady.}''
% % P7 mentioned, ``Some girls around me also have a Western style, and I think they are quite beautiful and healthily confident.'' P6 said they did not reject the V-shaped face, noting, ``Both V-shaped and round faces can be beautiful.'' 
% Such attitudes, to some extent, reflect the aesthetic diversity that results from the ongoing cultural exchange between the East and the West. Despite this, as shown in \autoref{sssec:ideal}, the indigenous standards of beauty still dominate, and a majority of participants expressed disapproval and discomfort with the Westernization of East Asian women. This suggests that, even amidst globalization, there remains significant tension between Eastern and Western values.

% Similar situations exist in reality. For example, in the current cosmetics markets of China, Japan, and Korea, \x{skin whitening remains a highly mainstream demand among woman users, while tanning is very niche~\cite{}.}

% While some East Asian users might accept the Westernized depictions of East Asian women generated by image-producing AI, such portrayals still risk evoking feelings of cultural invasion and provoke negative reactions such as denial, skepticism, and resistance.

\textbf{Sexualization \& feminization.}
In our study, many participants reported uneasiness of seeing Easat Asian women portrayed with revealing clothing.
For example, $P12_{CN,SD}$ was surprised by the output image of Stable Diffusion: ``\textit{I didn't expect it to be a girl like this, who gives off the impression of a woman of the street, and her clothing is also quite revealing.}''
% In another AI-generated image, the woman wears a flesh-colored, tight outfit, and $P15_{CN,SD}$ said, ``\textit{the image that comes to mind is that of a foot massage therapist (a role often associated with sensual or erotic services).}'' 
When an image of a woman in a deep V-neck outfit was generated, $P22_{KR,DE}$ said, ``\textit{We're not fond of that style; its neckline is a bit too low.}'' $P25_{JP,DE}$ commented that they usually ``\textit{wouldn't wear a V-neck that plunges so low, and a tight skirt; it doesn't seem to represent East Asia well.}''
Participants also noted several features that emphasize women's passivity and obedience, such as porcelain skin, dull eyes, and portrayals of them as consistently thin, young, and with long hair.
For example, 
% $P8_{CN,SD}$ reported that ``\textit{I was given a round-faced, youthful beauty. She is beautiful indeed, but it does not align with my impression of the typical East Asian mature beauty.}''
$P5_{CN,MJ}$ attempted to draw a middle-aged Chinese woman using the three models but was frustrated to find that ``\textit{these models seem not to understand how women can maintain their beauty as they age. They either ignore my prompts and insist on drawing young girls, or they just awkwardly add a few wrinkles.}''
$P1_{CN,DE}$ wanted to draw a woman with a plump figure, which used to be a sign of wealth and health in East Asia. She tried keywords such as ``plump'', ``fat'', ``chubby'', as well as more specific weight descriptors like ``60kg'', ``80kg''. However, the model continued to produce images of slim women.
% East Asian culture once symbolized a plump figure as a sign of wealth and health, and also celebrated the beauty of mature women~\cite{han2003body,cashin2024evolution}, but when our participants tried to draw such woman images in AI, they usually failed to do so. 
% $P8_{CN,SD}$ found that in some images,``\textit{the eyes do not seem to focus well, giving an impression of insecurity.}'' 
$P9_{CN,SD}$ found that in some images, ``\textit{the eyes seem lifeless, and to put it bluntly, it's as if the eyes suggest a lack of intelligence.}''
$P6_{CN,SD}$ said, ``\textit{the gaze is very rigid, which fits the discriminatory stereotype of East Asians by Westerners - thinking we are not wise enough, more docile, and obedient.}''
% $P18_{CN,SD}$: The people are all particularly emaciated
Moreover, while participants generally prefer fair skin, they disliked the doll-like, artificial paleness (\eg $P13_{CN,SD}$: ``\textit{It might perceive East Asian women as somewhat petite and delicate, or as having a kind of pitiable charm}''). Several participants thought depicting East Asian women uniformly with long hair is biased ($P4_{CN,DE}$: ``\textit{All have straight, black, long hair... If everyone is generated to be like this, I think it is a bias.''})
% After generating a series of images, $P13_{CN,SD}$ commented that the model ``\textit{might perceive East Asian women as having a certain quaint and delicate, or a kind of pitiable charm.}'' 
The aforementioned features closely resemble stereotypical archetypes once identified in Western narratives, such as the China Doll and the Geisha Girl~\cite{yamamoto2000visible}.
Besides, we also observed a few cases related to another sexualized archetype, namely the Dragon Lady, a mysterious and dangerous woman who allures men. Our participants primarily perceived this pattern through the women's harsh gaze (\eg $P8_{CN,DE}$: ``\textit{It feels a bit too severe, I mean her temperament}''; 
$P20_{CN,MJ}$: ``\textit{The gaze is quite fierce, not nice.}'')
% $P5_{CN,MJ}$: It lacks a sense of liveliness, giving the impression that the overall aura is rather downcast (lifeless eyes)
% These two types of gazes may seem distinct at first glance, but they actually share a similar root. 
% As Yamamoto~\cite{yamamoto2000visible} stated, ``whether delicately or dangerously sexualized'', they both represent sexualized views of East Asian women from a male perspective. Currently, these stereotypes that used to be pervasive in Western narratives, continue to be perpetuated in AI.

% In China's first anthology of poetry, the Book of Songs, there is a poem that praises the tall and plump wife of Duke Wei, who was admired at the time. 
% The term "tall person" referred to someone of a large and white complexion, which was considered beautiful at the time. 
% The poem also describes her as having a "charming smile" and "beautiful eyes," emphasizing dynamic charm rather than rigid standards. 

% In traditional Chinese social and family education, clothing codes of conduct are regarded as one of the important contents of self-cultivation, which has long affected the Chinese people’s hobbies and lives. Chinese pay more attention to the content of morality and ethics in terms of dress, and use clothing to cover up human beauty and not expose, so as to meet the moral requirements of Confucianism. Traditional Chinese clothing pays attention to the beauty of people’s spirit, temperament and charm, and does not emphasize highlighting the body but cares about the inside

\textbf{Overuse or misuse of cultural symbols.}
% Exoticism
The third reported pattern is the overuse or misuse of cultural symbols (\autoref{fig:examples}, third row). This includes the awkward and deliberate integration of ancient symbols (\eg traditional headdresses, earrings, clothing patterns, umbrellas, lanterns), as well as positioning East Asian women in front of traditional architecture like temples and gardens. For instance, $P1_{CN,MJ}$ noted that Midjourney ``\textit{generates images of ancient clothing, which feels like a stereotype. When I asked for modern clothes, it still provided me with ancient costumes.}'' $P8_{CN,DE}$ questioned: ``\textit{Why do they have to wear that kind of traditional clothing? I find it quite strange.}'' 
% $P21_{JP,MJ}$, who drew a Japanese woman who is working, complained, ``\textit{Japanese company employees do not wear kimonos.}'' 
$P26_{JP,MJ}$ generated an image of a Japanese mother playing with her children but was depicted in a kimono, which she found odd: ``\textit{It is weird, as mothers generally do not wear kimonos to parks.}'' $P27_{KR,DE}$ drew a Korean girl shopping, but the generated background was a vintage shop selling traditional Korean dress: ``\textit{It is not a typical place we would go to.}''
The second perceived bias is the confounding use of cultural elements, which fails to distinguish between the cultural differences within East Asia. 
% This confusion sometimes extends to broader Eastern regions, such as Southeast Asia and South Asia, resulting in a mismatch of symbols from different countries or cultural systems. 
For instance, $P21_{JP,DE}$ reported a case where she attempted to draw a Japanese woman, but the output ``\textit{looks very much like my perception of women in Chinese period dramas, but I wrote that she should be Japanese.}'' $P25_{JP,MJ}$, intending to draw a Japanese individual as well, said, ``\textit{Seeing the costume, I felt momentarily bewildered. This should be a Hanfu dress from China, what's the situation?}'' 
$P22_{KR,MJ}$ complained that ``\textit{South Korea never had umbrellas like this; these are entirely Japanese.}''
% $P10_{CN,MJ}$ asked Midjourney to draw a beautiful woman from East Asia and found: ``\textit{It indeed incorporates a bit of everything, but you can't say it's distinctly Chinese, Japanese, or Korean.}''
The third bias is the wrong semantics, that is, the application of symbols deviates from the actual situation. For example, $P23_{KR,DE}$ generated an image depicting a woman eating, but observed, ``\textit{When consuming this type of meal in South Korea, a spoon is typically used, yet the image showed chopsticks.}'' $P18_{CN,DE}$ requested an AI-generated image of a beautiful woman from northern China but found that the background elements are all from the scenery of southern provinces: ``\textit{There's no such terrain in the north.}''
% when a classical beauty was depicted using a modern laptop, P4 remarked, "I have no idea where this image should be used." P11 also noted, "The generated clothing is strange; in reality, no one would wear something like that."
The way AI mishandles East Asian cultural symbols echoes what Ziff and Rao~\cite{ziff1997introduction} termed cultural appropriation: the incorporation of East Asian cultural symbols by AI is not based on genuine understanding but is primarily used to create an exotic atmosphere. These symbols are often assembled in an arbitrary and decontextualized manner, resulting in portrayals that seem unnatural and disjointed.

\textbf{Racial stereotype.}
Several participants spotted the classic racial stereotype of slanted eyes. $P25_{JP,MJ}$ found that ``\textit{the eyes seem squinty, a typical slanted eyes.}''
$P8_{CN,DE}$ commented, ``\textit{Monolid eyes, upturned eye shapes; it may not rise to the level of racial discrimination, but it could indeed be a stereotype.}'' 
% Similarly, $P28_{JP,SD}$ noted, ``\textit{The eyes are positioned upwards here, which is a bit of a stereotype.''}. 
% $P5_{CN,MJ}$ The eyes are relatively narrow and elongated, and, hmm, the eyebrows are also quite narrow and elongated.
% However, participants still tended to report this bias more intuitively and expressed greater anger compared to other bias patterns (\eg when viewing such images, participants often first notice the shape of the eyes). 
In East Asia, slanted eyes are not merely a biological trait; they are also associated with the collective trauma stemming from East Asia's history of being discriminated against and exploited by the West. However, this pattern's occurrence in AI-generated images is relatively low compared to past patterns. We speculate this is due to the strict ethical reviews on many AI platforms. Yet, our participants were notably sensitive to and immediately noticed the shape of slanted eyes in images once it emerged.
Although some participants acknowledged that certain East Asians do have such eye shapes, they hoped that AI would not emphasize these eyes or would draw more aesthetically pleasing eyes (\eg $P22_{KR,DE}$: ``\textit{I feel uncomfortable seeing this; or, let's say, at least I don't find them beautiful. I think it overly emphasizes the upward slant of the eyes of East Asians.}'' $P26_{JP,SD}$: ``\textit{In my life, most people do not have this eye shape.}'')
% While these features are explicitly presented in AI-generated images, they can still manifest occasionally or in more subtle ways (\eg through eye makeup).

% Another interesting finding, as introduced earlier, is that a considerable number of participants reported biases related to slanted eyes were achieved through eye makeup. \w{For example, xxx}. In other words, even without generating an obviously elongated, upward-slanted eye shape, the upward stroke of the makeup can also trigger sensitive points for East Asians and be perceived as 'essentially' implying slanted eyes (this is very similar to the incident where a Dior advertisement faced backlash from East Asia due to the model's eye makeup and gesture, as shown in \autoref{fig:intro} A). 

%Asian and Asian American women are victims of the “double jeopardy” of racism and sexism. -- Asian Eyes: Body Image and Eating Disorders of Asian and Asian American Women

\subsubsection{Observations}
\label{sssec:observation}

As shown in \autoref{tab:taxonomy}, the distribution of identified biases varied across models. Of the \nbias biases, 9 (50.00\%) were most frequently observed in DALL-E, 6 (33.33\%) in Stable Diffusion, 2 (11.11\%) in Midjourney, and 1 (5.56\%) was equally observed in DALL-E and Stable Diffusion. 
The three models also exhibited different distributions across the four bias patterns. DALL-E was most frequently associated with \textit{Westernization} biases. Midjourney was most frequently reported for biases related to the \textit{overuse/misuse of cultural symbols}. For biases related to \textit{sexualization \& feminization} and \textit{racial stereotypes}, Stable Diffusion was the most frequently reported model. Overall, participants expressed the highest satisfaction with Midjourney (mean satisfaction score = 72.2 out of 100), followed by DALL-E (mean satisfaction score = 62.8). Despite having fewer dominating biases than DALL-E, Stable Diffusion often portrayed East Asian women in ways that were overly revealing, ingratiating, and with slanted eyes, which led to strong disapproval from participants, resulting in the lowest satisfaction rating (mean satisfaction score = 52.3).

% In terms of overall frequency, Stable Diffusion was the model for which participants reported biases the most frequently (N = xx), followed by DALL-E (N = xx) and Midjourney (N = xx). 

\section{Conclusion}

In this work, we explore the portrayal of East Asian women in three image-generating AI models (DALL-E, Midjourney, Stable Diffusion) through a user-centered auditing study. As a result, we identified a range of perceived biases in AI-generated images of East Asian women. While racial stereotypes were relatively infrequent, biases manifested in various other ways, causing the AI-generated images to diverge from the ideals and expectations of East Asian users. Besides, the manifestation of these issues varied across different models: DALL-E exhibited the most notable Westernization of East Asian women, Stable Diffusion showed the most significant sexualization, and Midjourney had considerable issues with the misuse of cultural symbols. Based on these findings, we advocate for more respectful and inclusive representation in future models.

\begin{acks}
This work was supported by the National Natural Science Foundation of China 62402121, Shanghai Chenguang Program, and Research and Innovation Projects from the School of Journalism at Fudan University. We thank all the participants in this study and the reviewers who gave us constructive feedback.
\end{acks}

\bibliographystyle{ACM-Reference-Format}
\bibliography{reference}
% \bibliography{reference2}

\end{document}